\documentclass[
superscriptaddress,
preprint,
showpacs,
 amsmath,amssymb,
 aps,
prb,
]{revtex4-1}

\usepackage{graphicx}
\usepackage{dcolumn}
\usepackage{bm}
\usepackage{hyperref}

\begin{document}

\title{Stabilization of an ambient pressure, collapsed tetragonal phase in CaFe$_2$As$_2$ and tuning of the orthorhombic / antiferromagnetic transition temperature by over 70 K by control of nano-precipitates. }

\author{S. Ran}
\author{S. L. Bud'ko}
\author{D. K. Pratt}
\author{A. Kreyssig}
\author {M. G. Kim}
\affiliation{Ames Laboratory, US DOE and Department of Physics and Astronomy, Iowa State University, Ames, Iowa 50011, USA}
\author{M. J. Kramer}
\affiliation{Ames Laboratory, US DOE and Department of Materials Science and Engineering, Iowa State University,  Ames, Iowa 50011, USA}
\author{D. H. Ryan}
\author{W. N. Rowan-Weetaluktuk}
\affiliation{Centre for the Physics of Materials and Physics Department, McGill University, Montreal, Quebec H3A 2T8, Canada}
\author{Y. Furukawa}
\author{B. Roy}
\author{A. I. Goldman}
\author{P. C. Canfield}
\affiliation{Ames Laboratory, US DOE and Department of Physics and Astronomy, Iowa State University, Ames, Iowa 50011, USA}

\date{\today}

\begin{abstract}
We have found a remarkably large response of the transition temperature of CaFe$_2$As$_2$ single crystals grown out of excess FeAs to annealing / quenching temperature.  Whereas crystals that are annealed at 400$^{\circ}$ C exhibit a first order phase transition from a high temperature tetragonal to a low temperature orthorhombic and antiferromagnetic state near 170 K, crystals that have been quenched from 960$^{\circ}$ C exhibit a transition from a high temperature tetragonal phase to a low temperature, non-magnetic, collapsed tetragonal phase below 100 K.  By use of temperature dependent electrical resistivity, magnetic susceptibility, X-ray diffraction, M\"ossbauer spectroscopy and nuclear magnetic resonance measurements we have been able to demonstrate that the transition temperature can be reduced in a monotonic fashion by varying the annealing / quenching temperature from 400 to 850$^{\circ}$ C with the low temperature state remaining antiferromagnetic for transition temperatures larger than 100 K and becoming collapsed tetragonal / non-magnetic for transition temperatures below 90 K.  This suppression of the orthorhombic / antiferromagnetic phase transition and its ultimate replacement with the collapsed tetragonal / non-magnetic phase is similar to what has been observed for CaFe$_2$As$_2$ under hydrostatic pressure. Transmission electron microscopy studies indicate that there is a temperature dependent, width of formation of CaFe$_2$As$_2$ with a decreasing amount of excess Fe and As being soluble in the single crystal at lower annealing temperatures.  For samples quenched from 960$^{\circ}$ C there is a fine (of order 10 nm), semi-uniform distribution of precipitate that can be associated with an average strain field whereas for samples annealed at 400$^{\circ}$ C the excess Fe and As form mesoscopic grains that induce little strain throughout the CaFe$_2$As$_2$ lattice.  
 
\end{abstract}


\maketitle

\section{Introduction}

CaFe$_2$As$_2$ manifests an extreme example of the coupled magnetic / structural phase transition that epitomizes the physics of the undoped parents of the FeAs-based superconductors.\cite{nin08a,can09a}  The strongly first order transition at ambient pressure from a high temperature, tetragonal, paramagnetic phase to a low temperature, orthorhombic, antiferromagnetic phase takes place near 170 K in single crystals grown out of Sn flux and manifests a hysterisis of several degrees as seen in thermodynamic, transport, and microscopic measurements.\cite{nin08a,can09a,gol08a,wug08a,ron08a}\\

CaFe$_2$As$_2$ is also the most pressure sensitive of the $A$Fe$_2$As$_2$ and 1111 compounds with its structural / magnetic phase transition being initially suppressed by over 100 K per GPa. \cite{can09a,tor08a,kre08a,gol09a,yuw09a,lee09a}   As pressure increases a non-magnetic, collapsed tetragonal phase that is stabilized by $\sim 0.3$ GPa intersects and terminates the lower pressure orthorhombic / antiferromagnetic phase line near 100 K and 0.4 GPa and rises to 300 K by $\sim 1.5$ GPa. \cite{tor08a,kre08a,gol09a}  In addition to this extreme pressure sensitivity, CaFe$_2$As$_2$ is also very sensitive to non-hydrostaticity.  \cite{can09a,kre08a,gol09a,yuw09a,lee09a,tor09a,pro10a}  If the pressure medium solidifies before the structural phase transitions, then the anisotropic changes in the unit cell lead to non-hydrostatic (by definition) stress, which in turn leads to dramatically broadened transitions and a structurally mixed phase sample in the 0.4 GPa pressure region.  This mixed phase includes a small amount of strain stabilized, high temperature tetragonal phase which superconducts at low temperature. \cite{can09a,tor08a,kre08a,gol09a,yuw09a,lee09a,pro10a,par08a}  The use of helium as a pressure medium allows for a minimization of these non-hydrostatic effects and has allowed for the determination of the $T - P$ phase diagram.  \cite{can09a,kre08a,gol09a,yuw09a}\\

Sn grown single crystals of CaFe$_2$As$_2$ are highly deformable and join the $R$Sb$_2$ and $R$Bi$_2$ compounds \cite{bud98a,pet02a} as rare examples of a malleable intermetallic compounds.  Single crystal plates can be bent, and, to some extent, even rolled by simply grasping with tweezers and applying minor torques across the sample length by pressing one end of the crystal downward on the surface of the lab bench, or microscope stage.  This malleability can lead to extreme broadening of features in ground samples, as were seen in early attempts at powder X-ray diffraction. \cite{nin08a} \\

CaFe$_2$As$_2$ samples were initially grown out of Sn and characterized in single crystal form. \cite{nin08a,ron08a}  Sn grown crystals are well formed, faceted plates that generally have planar dimensions of several mm and thicknesses between 0.1 and 0.5 mm. \cite{nin08a,can09a}  For measurements that require larger sample volumes pseudo-polycrystalline \cite{kre08a} or oriented single crystalline assemblies (see Figure 1 in Ref. \onlinecite{pra09a}) can be used.  Larger single crystals of CaFe$_2$As$_2$ have been grown out of ternary melts rich in FeAs. \cite{gol09a}  In order for these larger crystals to manifest a structural / magnetic phase transition similar to that seen in the smaller Sn grown crystal they were annealed at 500$^{\circ}$ C for 24 hours (a temperature similar to the decanting temperature of the Sn grown samples).  Without this annealing the larger, FeAs-grown, samples had dramatically suppressed transition temperatures.\\
	
Given recent observations of small shifts in the structural and magnetic transition temperatures of BaFe$_2$As$_2$ samples, and of the superconducting transition in doped BaFe$_2$As$_2$, as well as sharpenings of their signatures in thermodynamic and transport data, \cite{rot10a,gof10a} we undertook a systematic study of the effects of post growth thermal treatment of FeAs grown single crystals of CaFe$_2$As$_2$.  We have discovered that once again CaFe$_2$As$_2$ is the extreme case in the $A$Fe$_2$As$_2$ series, manifesting a surprisingly large suppression of the structural / magnetic transition temperature in “as grown” samples (nearly 50\%) that, even more remarkably, can be systematically changed from $\sim 170$ K to below 100 K with the lowest transition temperature samples having a transition into the non-magnetic, collapsed tetragonal state, but at ambient pressure.\\

In order to characterize and understand the effects of temperature treatment, as well as the nature of the low temperature state, we have performed a wide variety of thermodynamic, transport, microscopic, and spectroscopic measurements.  Temperature dependent electrical resistivity and magnetic susceptibility measurements were used to determine a transition temperature ($T^*$) – annealing temperature ($T_a$) phase diagram as well as identify similarities between the collapsed tetragonal phase and the low temperature state of FeAs-grown CaFe$_2$As$_2$ crystals quenched from temperatures between 850 and 960$^{\circ}$ C.  For annealing temperatures $ T_a \gtrsim 400$$^{\circ}$ C the $T^* - T_a$ phase diagram is found to be remarkably similar to the $T^*$-pressure ($P$) phase diagram, bringing up the question of what could the relationship between $T_a$ and $P$ be?  Temperature dependent single crystal X-ray diffraction measurements were then employed to unambiguously show that the crystallographic phase transition in as grown samples quenched from 960$^{\circ}$ C is one to a collapsed tetragonal state that is in qualitative as well as quantitative agreement with what is found for Sn-grown samples under applied pressures of $\sim 0.4$ GPa.  Temperature dependent M\"ossbauer spectroscopy measurements showed that the low temperature magnetic state of annealed FeAs-grown CaFe$_2$As$_2$ single crystals remains antiferromagnetic until the transition temperature is suppressed to below 100 K when the low temperature ground state becomes non-magnetic, a result confirmed by nuclear magnetic resonance (NMR) measurements.  Finally, transmission electron microscopy (TEM) measurements revealed that there is a small, temperature dependent width of formation for CaFe$_2$As$_2$, allowing for a solid solubility of excess Fe and As in the single crystals that decreases with temperature.  As the quenching temperature is reduced from 960$^{\circ}$ C to 400$^{\circ}$ C the initially fine precipitate coarsens, decreasing the degree of strain detected in the sample.

\section{Experimental Methods}

Single crystals of  CaFe$_2$As$_2$ were grown out of excess FeAs by rapidly cooling a melt of  CaFe$_4$As$_4$ from 1180$^{\circ}$ C to 1020$^{\circ}$ C over 3 hours, slowly cooling from 1020$^{\circ}$ C to 960$^{\circ}$ C over 35 hours and then decanting off the excess liquid, essentially quenching the samples from 960$^{\circ}$ C to room temperature.  These samples will be referred to as, "as grown samples".  Post growth, thermal treatments of samples involve the following variables:  annealing temperature, annealing time, and annealing environment.  Annealing environment refers to either (i) annealing a whole, unopened, decanted growth ampoule, or (ii) annealing individual crystals that have been picked from a growth and resealed in evacuated silica tubes.  For studies of the effects of annealing temperature, we seal several crystals into an evacuated silica tube and anneal for 24 hours in a furnace stabilized at the specified temperature.  The sample is placed into the hot furnace and, after annealing it is quenched to room temperature.  Longer time anneals (seven days) were used to prepare whole, unopened batches of samples.  In order to study the effects of annealing on FeAs grown samples with transition temperatures (and features) like the Sn-grown crystals, samples that had been annealed for a week at 400$^{\circ}$ C were subsequently sealed into an evacuated silica tube and annealed for 24 hours in a furnace stabilized at the specified temperature.  Although a detailed study of the annealing time dependence of sample changes will need to be done in the future, we found that, for example, at 450$^{\circ}$ C a one hour anneal is not enough to effect complete change, but anneals longer than 4 hours do not lead to any further significant changes in sample behavior;  at 800$^{\circ}$ C annealing appears to be completed in under 0.5 hour.  \\
 
Temperature dependent magnetization measurements were made in a  Quantum Design (QD) MPMS unit.  Temperature dependent electrical resistivity was measured in a four probe configuration, with Pt wires attached to the samples by Du Pont 4929N Ag-paint (cured at room temperature), in a QD PPMS unit.  Although normalized resistivity values are plotted, the resistivity values of samples did not vary outside of the uncertainty associated with a combination of geometric error (associated with measuring dimensions of the sample) and difficulties associate with sample exfoliation.  The average, room temperature resistivity of as grown, 700$^{\circ}$ C annealed and 400$^{\circ}$ C annealed samples was $3.75 \pm 0.75$ m$\Omega$ cm (a 20\% variation).\\     

In order to identify the nature of the structural transition in the as-grown  CaFe$_2$As$_2$ crystal (quenched from 960$^{\circ}$ C) and to determine the temperature dependence of the lattice parameters, high-energy X-ray diffraction measurements ($E = 99.62$ keV) using an area detector were performed on the 6-ID-D station in the MUCAT Sector at the Advanced Photon Source.  At this high energy, X-rays probe the bulk of a crystal rather than just the near-surface region and, by rocking the crystal about both the horizontal and vertical axes perpendicular to the incident X-ray beam, an extended range of a chosen reciprocal plane can be recorded. \cite{kre07a}  For the measurements, the horizontal angle, $\mu$ was scanned over a range of $\pm 3.6 ^{\circ}$ for each value of the vertical angle, $\eta$, between $\pm 3.6 ^{\circ}$ with a step size of $0.4 ^{\circ}$.  The two-dimensional scattering patterns were measured by a MAR345 image-plate positioned 1503 mm behind the sample.  The crystal was mounted on the cold finger of a closed-cycle refrigerator surrounded by a beryllium heat shield and vacuum containment.  Additionally, the crystal was mounted such that there was access to $(h k 0)$, $(h 0 l)$, and $(h h l)$ reciprocal lattice planes.  The $(0 0 8)$ and $(2 2 0)$ peaks were fit for lattice parameter determination and, for these measurements, the total exposure time for each frame was 383 s.  \\ 

The M\"ossbauer absorbers were prepared by attaching several single crystal plates to a 12 mm diameter disc of 100 $\mu$m thick Kapton foil using GE-7031 varnish. The spaces between the crystals were filled with a radio-opaque paint prepared by mixing 1-5 $\mu$m tungsten powder (obtained from Alpha-Aesar) with diluted GE-7031 varnish. The absence of gaps in the completed mosaic was confirmed first visually and then by looking for transmission of the 6.4 keV Fe-K$_{\alpha}$ X-ray from the $^{57}$Co M\"ossbauer source. In the configuration used, the crystalline $c$ -axis was parallel to the M\"ossbauer $\gamma$-beam.\\

The M\"ossbauer spectra were collected on a conventional spectrometer using a 50 mCi $^{57}$Co/Rh source mounted on an electromechanical drive operated in constant acceleration mode. The spectrometer was calibrated against $\alpha$-Fe metal at room temperature. Temperatures down to 5 K were obtained using a vibration-isolated closed-cycle refrigerator with the sample in a partial pressure of helium to ensure thermal uniformity. Spectra were fitted using a conventional, non-linear, least-squares minimization routine to a sum of equal-width Lorentzian lines. The line positions for the magnetic sextets observed in the ordered state were calculated assuming first-order perturbation in order to combine the effects of the magnetic hyperfine field and the electric field gradient. As the samples were oriented mosaics rather than powders, the line intensities were constrained to be in the ratio $ 3:R:1:1:R:3$ (following the conventional practice of labeling the lines from negative to positive velocity) \cite{gre71a} with the intensities of the two $\Delta m_I  =0$ lines being variable (“$R$”) to allow for the expected magnetic texture. $R = 0$ would correspond to the moments being parallel to the M\"ossbauer $\gamma$-beam, whereas $R = 4$ indicates that the moments are perpendicular to the beam.\\

Nuclear magnetic resonance (NMR) measurements were carried out on $^{75}$As ($I = 3/2$; $\gamma/2 \pi = 7.2919$ MHz/T) by using a homemade phase-coherent spin-echo pulse spectrometer, to investigate the magnetic and electronic properties of differently treated CaFe$_2$As$_2$ crystals from a microscopic point of view. $^{75}$As-NMR spectra at a resonance frequency of 51 MHz were obtained by sweeping the magnetic field. \\

TEM samples were prepared by mechanically polishing the single crystal to $\sim 10$ $\mu$m thick along the c-axis and then ion milling to perforation using 3 keV $\sim 18 ^{\circ}$ incident angle and following up with 30 min at 500 eV at $10^{\circ}$ to remove milling damage.  All milling was performed using a liquid N$_2$ cooled stage (sample $T \sim 120$ K).  Samples were analyzed using a Philips CM30 TEM operated at 300 keV.  Energy dispersive spectroscopy (EDS) and selective area diffraction patterns (SADP) were also performed on the samples in the TEM.\\

\section{Data Presentation}

\begin{figure}
\begin{center}
\includegraphics[angle=0,width=100mm]{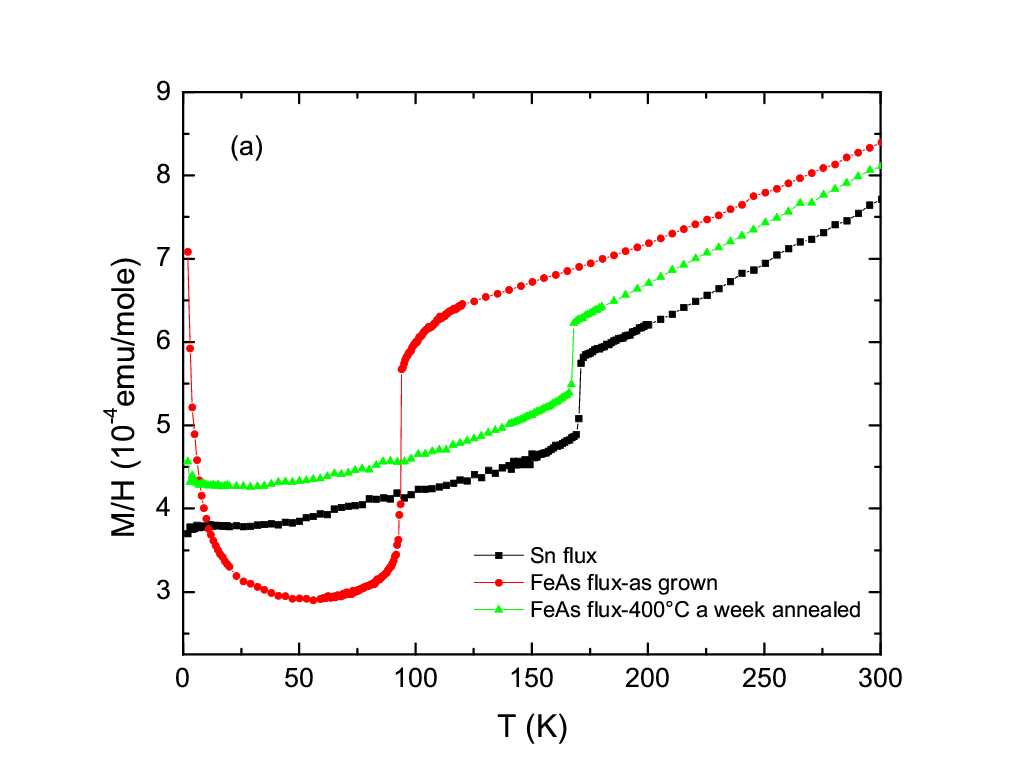}
\includegraphics[angle=0,width=100mm]{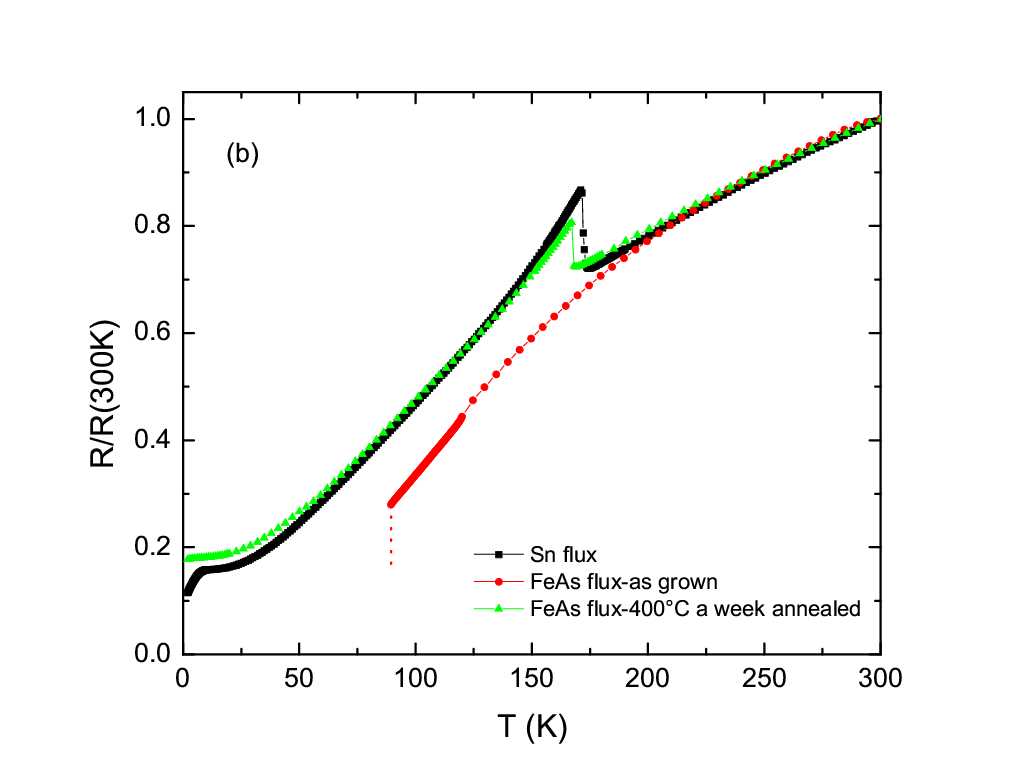}
\end{center}
\caption{\label{F1} (Color online)  Temperature dependent (a) magnetic susceptibility and (b) normalized electrical resistivity of CaFe$_2$As$_2$ for three differently prepared single crystals:  squares - Sn grown; circles - as grown (quenched from 960$^{\circ}$ C) from FeAs;  triangles - FeAs grown, annealed for a week at 400$^{\circ}$ C.  Note:  when the as grown sample from FeAs melt is cooled below the transition temperature near 90 K it shattered, making further lower temperature resistivity measurements impossible.
 }
\end{figure}

Figure \ref{F1} presents the resistivity and magnetic susceptibility for CaFe$_2$As$_2$ single crystals grown out of Sn and for CaFe$_2$As$_2$ single crystals grown out of excess FeAs.  Two data sets are shown for FeAs grown crystals:  one data set shows measurements on an as grown crystal that was decanted at, and quenched from, 960$^{\circ}$ C; the other data set shows measurements on a sample from a batch that was subsequently annealed at 400$^{\circ}$ C for a week.  The Sn-grown single crystal and the FeAs grown sample that has been annealed at 400$^{\circ}$ C are quite similar, both manifesting similar, modest increases in resistivity and decreases in susceptibility associated with the phase transition near 170 K. \cite{nin08a,can09a} On the other hand, the FeAs sample that was quenched from 960$^{\circ}$ C shows a significantly larger, very sharp drop in magnetization occurring well below 100 K.  The electrical resistivity also drops discontinuously at this temperature, associated with the sample suddenly undergoing a violent structural phase transition that often (usually) leads to shattering along the length and width of the bar, as well as loss of contacts.\\  

In addition to the quantitative differences shown in Fig. \ref{F1}, there is a qualitative difference between the as grown, CaFe$_2$As$_2$ single crystals from FeAs solution and the single crystals grown from Sn.  Whereas the Sn-grown single crystals are malleable and can easily be bent and deformed, the crystals quenched from a 960$^{\circ}$ C FeAs solution are brittle and tend to shatter if bending is attempted.  The FeAs grown crystals that have been annealed at 400$^{\circ}$ C, however, recover some of the malleability of the Sn grown ones and can deform a little without shattering.\\

	Given the dramatic difference in transition temperature, as well as the different signatures of the transition in resistivity and magnetization, several questions arise. Among them we consider:  (i) what is the nature of the phase transition in the as grown sample and (ii) can the transition in annealed samples be varied from near 170 K to below 100 K in a systematic manner?  We will address the latter question first and return to the former after the creation of a transition temperature ($T^*$) - annealing temperature ($T_a$) phase diagram. \\

\begin{figure}
\begin{center}
\includegraphics[angle=0,width=100mm]{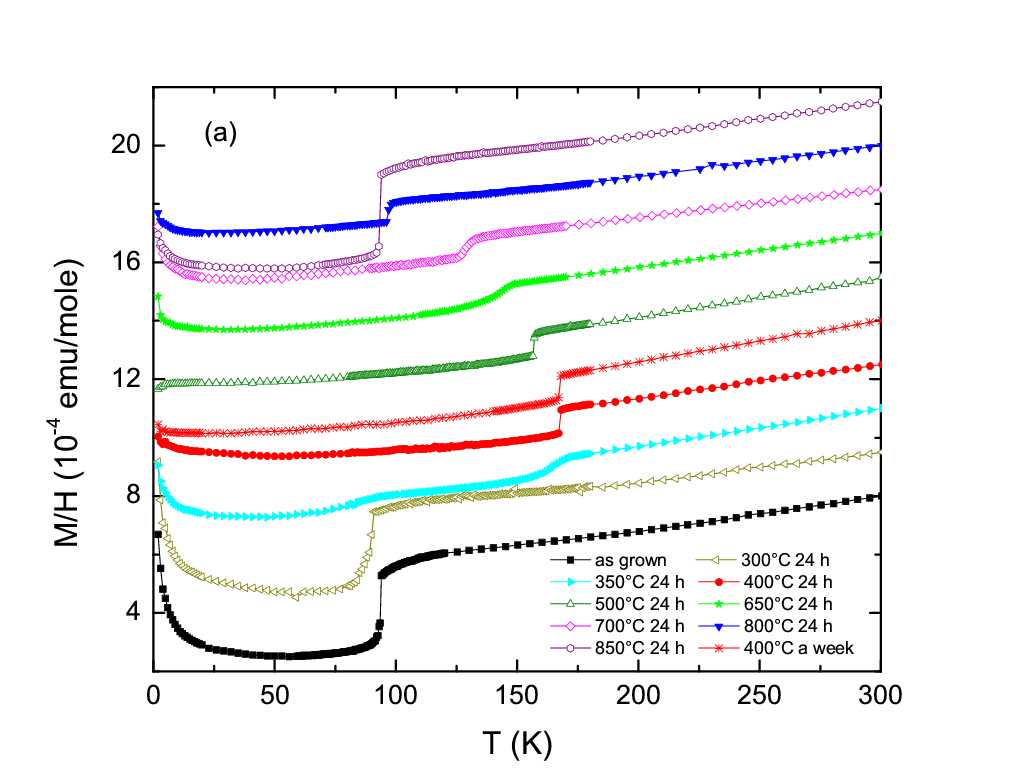}
\includegraphics[angle=0,width=100mm]{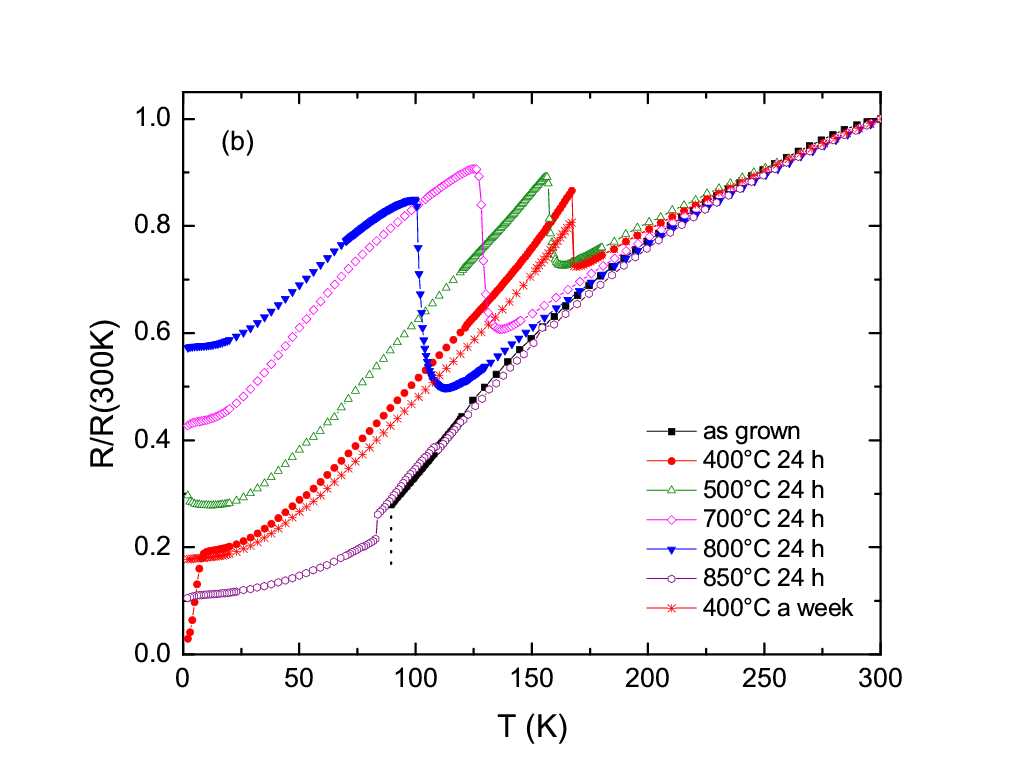}
\end{center}
\caption{\label{F2} (Color online)  Temperature dependent magnetic susceptibility and normalized electrical resistivity of as grown CaFe$_2$As$_2$ single crystals annealed for 24 hours at temperature, $T_a$.   Susceptibility data in (a) have been offset from each other by an integer multiple of $1.5 \cdot 10^{-4}$ emu/mole for clarity.  Data for a one week anneal of a whole batch at 400$^{\circ}$ C is shown for comparison.  The resistivity data (b) for the as grown sample could not be measured below the transition temperature due to sample breakage, but for the sample annealed at 850$^{\circ}$ C resistivity could be measured through the transition. }
\end{figure}

	In order to assess the extent to which the 170 K phase transition that occurs in Sn-grown, as well as annealed FeAs-grown, samples of CaFe$_2$As$_2$ can be systematically shifted down to below 100 K we measured the temperature dependent susceptibility and resistivity of as grown samples that were annealed for 24 hours at temperatures ranging from 250 to 850$^{\circ}$ C.  Figure \ref{F2} presents  magnetic susceptibility and resistivity data for representative annealing temperatures.  The decrease in susceptibility (or increase in resistivity) can be shifted down in temperature by choosing an appropriate annealing temperature between 400 and 800$^{\circ}$ C.  For annealing between these temperatures the transitions, particularly as seen in the resistivity data, remain quite sharp and shift in a systematic manner.  Whereas the size of the jump in the magnetization remains fairly constant in the samples annealed in this temperature region, there is a monotonic increase in the magnitude of the increase in the resistivity (see Fig. \ref{F7} below).  \\

\begin{figure}
\begin{center}
\includegraphics[angle=0,width=100mm]{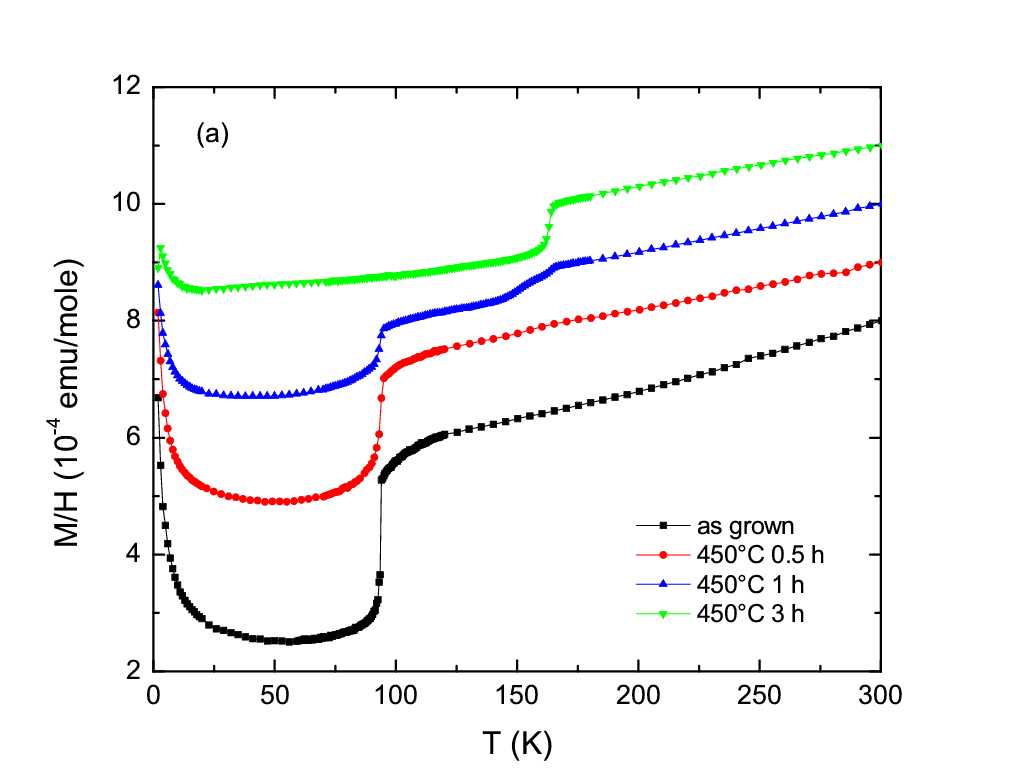}
\includegraphics[angle=0,width=100mm]{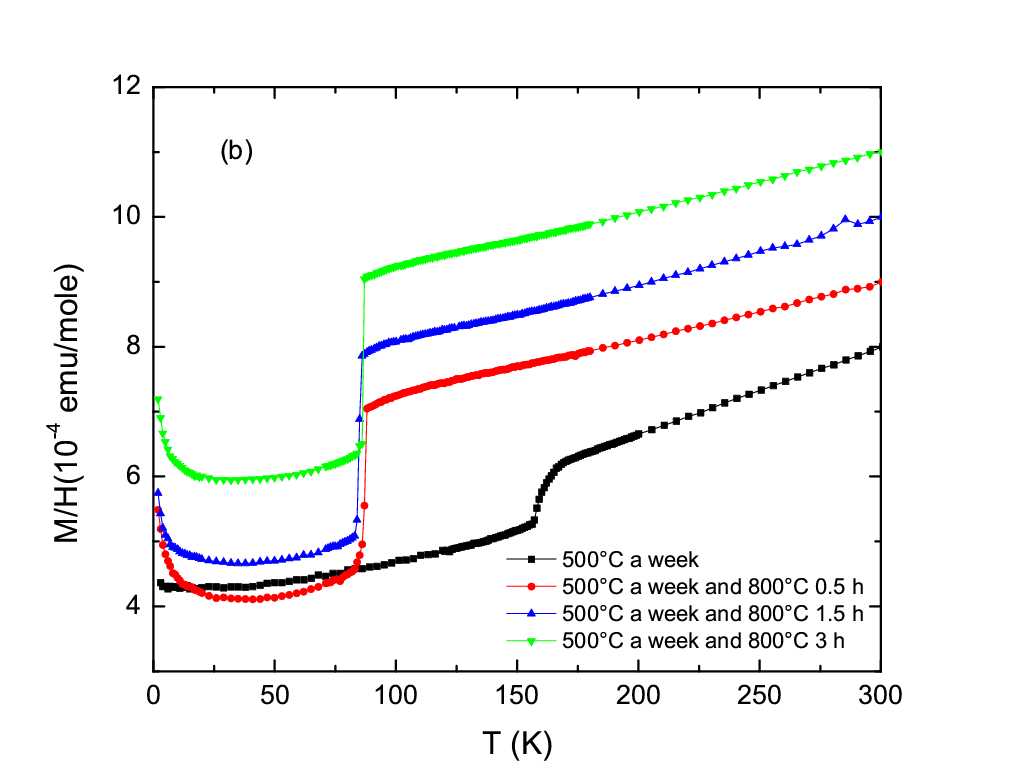}
\end{center}
\caption{\label{F3} (Color online)  Temperature dependent magnetic susceptibility of (a) as grown CaFe$_2$As$_2$ single crystals annealed at 450$^{\circ}$ C for representative times and (b) as grown CaAs$_2$Fe$_2$ single crystals that have been annealed for a week at 500$^{\circ}$ C and then annealed at 800$^{\circ}$ C for representative times. Data in both panels have been offset from each other by an integer multiple of $1 \cdot 10^{-4}$ emu/mole for clarity.  }
\end{figure}

	Such a clear temperature dependence of the effects of annealing, over such a wide temperature range, begs the question of what the annealing time dependence of these effects is.  In other cases of clear annealing effects, both time and temperature cuts through phase space are needed to establish unambiguous annealing protocols.\cite{mia02a}  In Fig. \ref{F3} we show the evolution of the magnetic susceptibility for different annealing times.  At 450$^{\circ}$ C, 0.5 hr is insufficient time to effect any significant change;  1.0 hr leads to split, broadened features with drops in susceptibility below both 170 and 100 K; 3.0 hr leads to a single, sharp feature near 170 K, comparable to what is seen for 24 hour anneals.  This progression shows that for 450$^{\circ}$ C, 24 hours is longer than the salient time scale for annealing.  As would be expected, for higher temperatures the salient time scale is even shorter.  In Fig. \ref{F3}(b) samples from a batch that had been annealed for a week at 500$^{\circ}$ C, with a transition temperature above 150 K, were annealed at 800$^{\circ}$ C for representative times.  As can be seen, even a 0.5 hour anneal causes the sample to behave in a manner similar to the as grown (quenched from 960$^{\circ}$ C) samples.\\

	Figure \ref{F2} also demonstrates that 24 hour anneals at temperatures of 300$^{\circ}$ C or lower temperatures do not change the temperature dependence of the as grown samples.  The data from the sample annealed at 350$^{\circ}$ C for 24 hours shows somewhat broadened drops in susceptibility near both 170 and 100 K, similar to what was seen for a 1.0 hr anneal at 450$^{\circ}$ C (Fig. \ref{F3}a), indicating that at 350$^{\circ}$ C 24 hours is comparable to, but less than the salient time scale.  Although longer annealing times for $T \lesssim 350$$^{\circ}$ C may lead to a sharp, single transition near 170 K (as is seen for the 400 and 500$^{\circ}$ C, 24 hr anneals) the time needed to achieve this state is anticipated to become exponentially long.  The one other data point we can add to this is the fact that 20$^{\circ}$ C (room temperature) anneals approaching $10^4$ hours have not led to significant changes in behavior of as grown samples.  \\

A 24 hour anneal at 850$^{\circ}$ C does not significantly change the transition temperature from that measured for the as grown samples quenched from 960$^{\circ}$ C (perhaps not too surprisingly since 850$^{\circ}$ C is approaching the 960$^{\circ}$ C quench temperature); the resistivity data for this sample, though, can be collected below the transition temperature, showing that the low temperature state has a lower resistivity, leading to a downward jump in resistivity when cooling through the transition temperature. \\

\begin{figure}
\begin{center}
\includegraphics[angle=0,width=100mm]{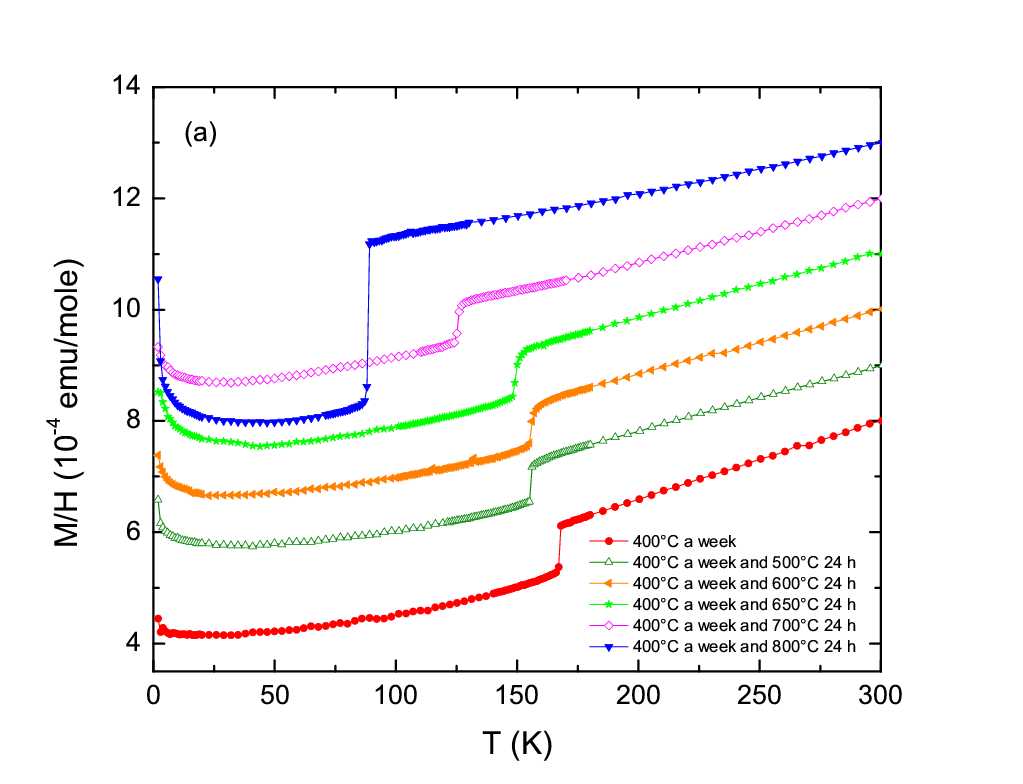}
\includegraphics[angle=0,width=100mm]{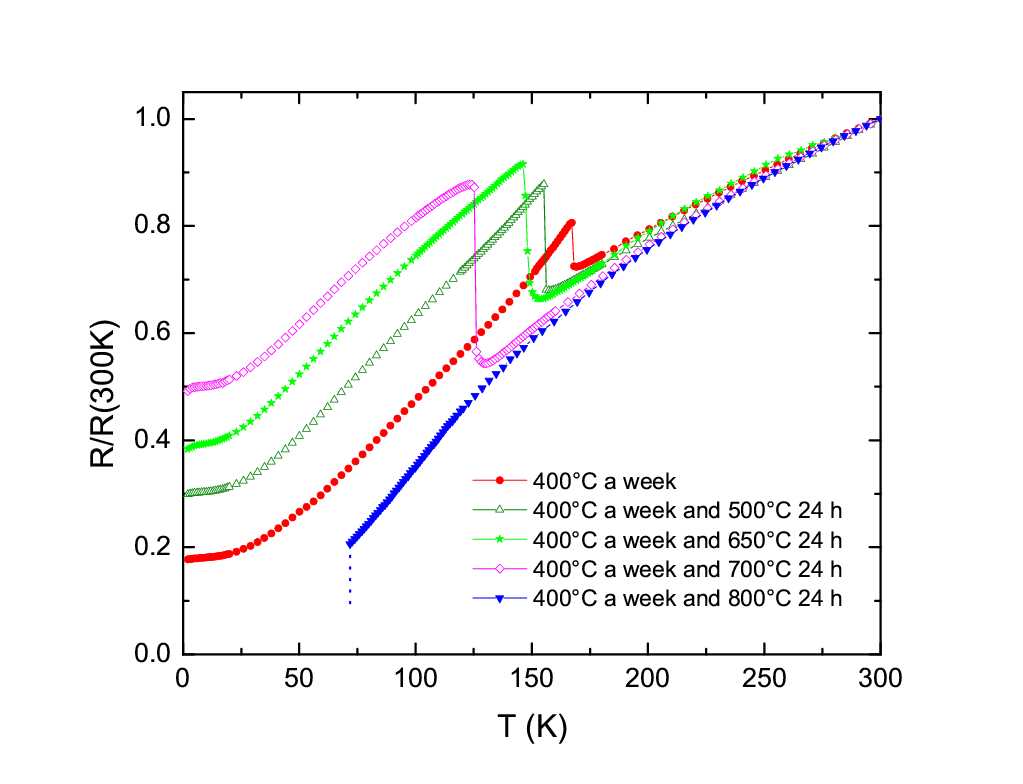}
\end{center}
\caption{\label{F4} (Color online)   Temperature dependent (a) magnetic susceptibility and (b) normalized electrical resistivity of as grown CaFe$_2$As$_2$ single crystals that were first annealed for a week at 400$^{\circ}$ C and then annealed for 24 hours at temperature, $T_a$.   Susceptibility data in (a) have been offset by an integer multiple of  $1 \cdot 10^{-4}$ emu/mole from each other for clarity.  
 }
\end{figure}

In order to see if similar changes in transition temperature could be induced by annealing samples that started with transitions near 170 K (i.e. started with transitions similar to those found in Sn grown CaFe$_2$As$_2$) we annealed an entire batch of crystals at 400$^{\circ}$ C for a week.  The resistivity and susceptibility data for these samples are also shown in Fig. \ref{F2} and are essentially the same as that found for the 24 hr anneal of individual crystals.  Single crystals from this, "400$^{\circ}$ C anneal for one week" batch were then separately sealed in silica ampoules and annealed for 24 hours at temperature ranging from 500$^{\circ}$ C to 800$^{\circ}$ C.  The temperature dependent resistivity and susceptibility for these samples are shown in Fig. \ref{F4}.  As was the case for the as grown samples, sharp features in both resistivity and susceptibility systematically shift to lower temperature when the sample is annealed at higher temperature.  The sample annealed at 800$^{\circ}$ C shows the larger drop in susceptibility and broke on cooling through its transition, making it appear to be similar to the as grown, quenched from 960$^{\circ}$ C, samples.  \\

Figure \ref{F5} presents the transition temperature -– annealing temperature, $T^*  - T_a$, plot.  Figure \ref{F6} illustrates how values for $T^*$, as well as the error bars, were inferred from the resistivity and susceptibility data.  As can be seen in Figs. \ref{F2} and \ref{F4}, for $T_a > 400$$^{\circ}$ C, there is a systematic progression of fairly sharp transitions downward for increasing $T_a$.  Figure \ref{F5} illustrates that, (i) there is some scatter in $T^*$ for a given $T_a$, but (ii) that there is also a fairly well defined suppression of $T^*$ with increasing $T_a$, e.g. a 400$^{\circ}$ C anneal gives a very different transition temperature from a 700$^{\circ}$ C anneal, which is itself different from a 850$^{\circ}$ C anneal or the as grown sample. In addition, annealing at a given $T_a$ leads to a $T^*$ value, regardless of whether the sample starts from a 170 K or $\sim 90$ K transition state, i.e. this final anneal determines $T^*$ regardless of sample history. \\

	The $T^* - T_a$ phase diagram presented in Fig. \ref{F5} shows that CaFe$_2$As$_2$ grown out of FeAs can have the temperature of its phase transition modified in an essentially continuous manner from near 170 K to below 100 K.  For transitions with $T^*$ between 170 K and 100 K the magnetic signature of the transition is essentially unchanged and the resistive signature evolves gradually with the jump in resistivity, $\Delta \rho$, becoming larger as $T^*$ decreases.  For the lowest $T^*$ values, below 100 K, there is a significantly larger drop in susceptibility and, when it can be measured, the jump in resistivity is downward on cooling rather than upward.  These observations are quantified in Fig. \ref{F7}.  As discussed in the experimental methods section, above, the room temperature resistivity of as grown samples as well as those annealed at 400 and 700$^{\circ}$ C for a week all have room temperature resistivity values of $3.75 \pm 0.75$ m$\Omega$-cm.  This invariance, within experimental resolution, allows for conversion of these jumps to absolute resistivity as needed. \\

	The evolution of the transition temperature (Fig. \ref{F5}) as well as the evolution of the resistive and magnetic signature of the phase transition (Figs. \ref{F1} - \ref{F4}) make it plausible that for 170 K $> T^* \gtrsim 100$ K the transition is similar to that seen in Sn-grown CaFe$_2$As$_2$:  a transition from a high temperature tetragonal, paramagnetic state to a lower temperature orthorhombic, antiferromagnetic state.  On the other hand, dramatic changes in the resistive and magnetic signature associated with the as grown sample as well as samples annealed near 850$^{\circ}$ C are consistent with our current understanding of the collapsed tetragonal phase, a phase that up until this time was associated with CaFe$_2$As$_2$ under pressures of 0.35 GPa or higher.  The change in the resistive signature of the transition from a sharp increase to a sharp decrease, as well as the sudden increase in the size of the drop in susceptibility upon cooling are very similar to the changes seen in Ref. \onlinecite{yuw09a} under hydrostatic pressure applied with helium.  For that matter, the basic phase diagram proposed in Refs \onlinecite{gol09a,yuw09a} is remarkably similar to the $T_a > 400$$^{\circ}$ C part of the $T^* - T_a$ phase diagram presented in Fig. \ref{F5} with $T_a$ playing the role of pressure, or more precisely stated somehow parameterizing the amount of stress in the sample.  This similarity can be seen in Fig. \ref{F5}b which directly plots $T^*$ as a function of $P$ and $T_a$.  For annealing temperatures that allow for the achievement of equilibrium by 24 hours (i.e. 400$^{\circ}$ C or greater) there is a remarkable agreement between the effects of $T_a$ and $P$ on $T^*$, at least as long as the transition is from high temperature tetragonal to low temperature orthorhombic / antiferromagnetic.  Once the low temperature state is the non-magnetic, collapsed tetragonal phase the annealing temperature does not seem to affect $T^*$ in the same manner as $P$.\\

	Before progressing too much further, though, we need to examine (and, as will be shown, verify) several of the hypotheses outlined above.  First we should parameterize and understand the nature of the low temperature ground state in the as grown FeAs crystals, once that is done we can return to the question of what may be causing the systematic changes we see in the $T^* - T_a$ phase diagram.  \\

	The collapsed tetragonal, CT, phase was identified \cite{kre08a,gol09a} in CaFe$_2$As$_2$ by scattering measurements made on samples under hydrostatic pressure using He as a pressure medium as part of a comprehensive effort to better understand the details of the CaFe$_2$As$_2$ $T - P$ phase diagram. \cite{tor08a,kre08a,gol09a,yuw09a,tor09a}  As CaFe$_2$As$_2$ transforms from the high temperature tetragonal phase into the low temperature, collapsed tetragonal phase the $c$-lattice parameter changes from $\sim 11.6$ \AA ~ to $\sim 10.6$ \AA, a remarkably large,  $\sim 10$\% decrease while the $a$-lattice parameter increases by $\sim 2$\%, leading to an $\sim 4$\% decrease in the unit cell volume. \cite{kre08a}  In order to see if the as grown sample manifests such a striking change in lattice parameters, high-energy single-crystal X-ray diffraction data were collected as a function of temperature. Figure \ref{F8} displays the results of fits to these data to extract the lattice parameters [Figs. \ref{F8}(a) and (b)] as well as the unit cell volume [Fig. \ref{F8}(c)].   The data from the original Sn-flux grown samples, at an applied pressure of 0.63 GPa, \cite{kre08a} are also included in Fig. \ref{F8} for direct comparison. These data clearly indicate that, structurally, the as grown crystals of CaFe$_2$As$_2$ transform into a collapsed tetragonal phase below 100 K at ambient pressure.  In particular, the inset to Fig. \ref{F8}(b) shows the diffraction image of the $(2~2~0)$ Bragg reflection at 6 K, the base temperature of our measurement.  Within our resolution, no splitting of the peak is evident as would be expected for an orthorhombic unit cell.  Furthermore, we find that the temperature dependence of the lattice parameters and unit cell volume are consistent with what was observed for the pressure induced, collapsed tetragonal phase for $P = 0.63$ GPa where the tetragonal-to- CT phase transition has moved above 150 K.    \\

	Although the sharp drop in susceptibility certainly suggests that the non-magnetic phase is associated with the collapsed tetragonal state, it is prudent to examine the magnetic properties of this state more closely with microscopic measurement techniques. M\"ossbauer spectroscopy measurements were carried out on three FeAs grown samples at selected temperatures. Composite samples, with aligned $c$-axes, were made from single crystals from whole batches treated in the following manner:  as grown (quenched from 960$^{\circ}$ C), annealed for a week at 400$^{\circ}$ C, annealed for a week at 700$^{\circ}$ C.  As shown in Fig. \ref{F5}, these three annealing temperatures produce samples representative of the whole range of behavior observed.  Whereas similar sample masses were used for the two extreme samples, a smaller sample mass was used for the sample annealed at 700$^{\circ}$ C, leading to slightly poorer signal and statistics. \\

  	The spectra of the three samples taken at 295 K (Fig. \ref{F9}) are essentially indistinguishable, showing a barely resolved quadrupole split doublet with the high velocity line slightly more than twice the intensity of the lower velocity line in each case.  The location of the Fe atoms in the ThCr$_2$Si$_2$-type structure requires that the principal axis of the local electric field gradient tensor lies along the $c$-axis, so our oriented mosaic should yield an intensity ratio of 3:1,\cite{gre71a} rather than the 2.3(3):1, 2.0(2):1 and 2.4(2):1 observed here for the 400$^{\circ}$ C annealed, 700$^{\circ}$ C annealed and as grown samples respectively.  It is unlikely that the reduced ratio is due to misalignment of the crystal plates as the intensities in the magnetic patterns are consistent with almost perfect alignment. It is possible that a minor impurity is the source of the reduced ratio, however no identifiable impurity contribution was found in the residual patterns of the fitted spectra, except in the case of the 700$^{\circ}$ C annealed sample, where about 4\% of the total iron was found to be associated with a non-magnetic phase at 5 K. The quadrupole splittings for the as grown and 700$^{\circ}$ C annealed samples were slightly higher (+0.236(5) mm/s) than that in the 400$^{\circ}$ C annealed sample (+0.202(12) mm/s) probably reflecting more distorted local Fe environments.  The sign of the electric field gradient cannot normally be determined from a $^{57}$Fe M\"ossbauer spectrum of a powder sample but the combination of oriented ThCr$_2$Si$_2$-type single crystal samples with the observed intensity ratio makes the sign determination unambiguous. \\

Cooling to low temperatures makes the differences between the three samples strikingly obvious. The two annealed samples undergo sudden transitions near 170 K (annealed at 400$^{\circ}$ C) and 130 K (annealed at 700$^{\circ}$ C) and by 10 K the sample annealed at 400$^{\circ}$ C has developed a clear magnetic splitting of 10.03(3) T whereas the sample annealed at 700$^{\circ}$ C exhibits a slightly smaller hyperfine field of  9.51(3) T. Fitting the intensities of the two $\Delta m_I = 0$ lines (lines 2 and 5) in both magnetic patterns yields $R= 3.8(1)$, implying an almost perfect (better than $10^{\circ}$) alignment of the crystal $c$-axis with the $\gamma$-beam, \cite{gre71a} as expected from the construction of the sample, and confirming that the ordering direction of the iron moments in both ordered samples lies in the $ab$-plane. By contrast, the spectrum of the as grown sample is almost unchanged. The quadrupole splitting increases slightly, to 0.272(4) mm/s, and there is also a small increase in linewidth (from 0.143(4) mm/s to 0.170(3) mm/s), possibly reflecting some increased disorder or strain. The largest change appears in the line intensity ratio, which drops to 1.70(7):1 on cooling to 5 K, suggesting either a reduction in quality of the $c$-axis alignment or a tilting of the principal axis of the local electric field gradient away from the c-axis.  Remarkably, the original intensity ratio is recovered on warming back to 295 K, so the change is fully reversible. Visual inspection of the three mosaics following several thermal cycles between room temperature and the base temperature of the cryostat did not reveal any apparent damage. \\

The temperature dependence of the hyperfine field for the two annealed samples, shown in Fig. \ref{F10},  reveals that not only is the iron moment probably slightly smaller in the sample annealed at 700$^{\circ}$ C, but the temperature dependence of the hyperfine field ($B_{hf}$) is visibly stronger, consistent with a lower ordering temperature. Although it is not possible to determine the ordering temperature of the hypothetical, second order phase transition of an antiferromagnetically ordered, orthorhombic phase transforming into a  paramagnetic, orthorhombic phase directly, since in reality the sample undergoes a strong first order transition to the tetragonal phase on warming, we can make an estimate by fitting the observed temperature dependence of $B_{hf}$ to a Brillouin function as is shown in figure \ref{F10}. This procedure yields estimated ordering temperatures of 300(10) K for the sample annealed at 400$^{\circ}$ C, and 230(10) K for the sample annealed at 700$^{\circ}$ C, with the errors dominated by an uncertainty in the effective total angular momentum quantum number, $J$, used in the fits. We note that whereas the absolute values of the N\'eel temperatures did depend somewhat on $J$, the difference between the values of $T_N$ for the two samples did not, and was consistently 70(3) K. The magnetic ordering in the sample with the lower structural transition temperature is definitely weaker, involving slightly smaller iron moments. \\

NMR measurements were also carried out on the as grown and 400$^{\circ}$ C annealed samples.  Fig. \ref{F11}(a) shows $^{75}$As-NMR spectra at $T = 200$ K for two magnetic field directions of $H \| c$-axis and $H \| ab$-plane for the 400$^{\circ}$ C annealed crystal. The observed quadrupole-split NMR spectra are well reproduced by a simple nuclear spin Hamiltonian \cite{car77a}  $H = \gamma \hbar \vec{I} \cdot \vec{H}_{eff} + {h \nu_Q \over 6} [3 I_Z^2 - I(I+1)]$, where $H_{eff}$ is the effective field at the As site (summation of external field $H$ and the hyperfine field $H_{int}$), $h$ is Planck's constant and $\nu_Q$ is nuclear quadrupole frequency which is proportional to the electric field gradient (EFG) at the As site (an assymetric parameter of EFG is assumed to be zero for simplicity). The blue lines in the figure show simulated spectra calculated from the simple Hamiltonian. Below 160 K, each NMR line for $H \| c$-axis splits into two lines due to internal field $H_{int}$ (parallel or antiparallel to $H$) which is produced by the Fe spin ordered moment. A typical example of the split NMR lines for $H \| c$-axis is shown at the bottom of Fig. \ref{F11}(a). The spectrum is reproduced well by $H_{int} = 2.59$ T and $\nu_Q = 12.7$  MHz at $T = 50$ K. These values are in good agreement with previously reported values for $^{75}$As-NMR of single crystals ($T_N = 167$ K) grown out of Sn flux, \cite{bea09a} once again indicating that the sample annealed at 400$^{\circ}$ C is essentially the same as previously reported ones grown out of Sn. The temperature dependence of the $^{75}$As spin lattice relaxation rates ($1/T_1$ ) measured at center line for $H \| c$-axis is also in good agreement with previous work. \cite{bea09a}  \\
 
    Similar quadrupole-split NMR spectra are observed in the as-grown CaFe$_2$As$_2$ sample as shown in Fig. \ref{F11}(b), but the observed $\nu_Q \sim 18-18.5$ MHz at $T = 140-110$ K is larger than that in the annealed crystal. The $^{75}$As-NMR satellite linewidth, which reflects the distribution of EFG, is significantly larger than in the annealed sample, indicative of higher degree of inhomogeneity of the local As environment due to strains, defects, or lattice distortion in the as-grown sample.  Below the transition temperature, $T \sim 96$ K, no splitting of the NMR lines is observed (indicating that there is no antiferromagnetic order) but $\nu_Q$ is found to change dramatically: from $\sim 18$ MHz to $\sim 42$ MHz, as is shown at the bottom in Fig. \ref{F11}(b).  Such a drastic change of $\nu_Q$ (more than 230\%) can not be explained by thermal expansion of lattice (at most few \%) but is attributed to a structural phase transition.  The value $\nu_Q  \sim 42$ MHz is also confirmed by the observation of nuclear quadrupole resonance (NQR) spectrum at zero magnetic field at $T= 4.2$ K (Fig. \ref{F11}(c)). The peak position in the NQR spectrum for the as-grown sample is higher than the 25 and 30.4 MHz for the tetragonal and collapsed tetragonal phases respectively in CaFe$_2$As$_2$ under high pressure. \cite{kaw11a} The combination of no splitting of the NMR lines with the large shift in $\nu_Q$ are further confirmation that, for the as grown sample, there is only a structural phase transition without any magnetic phase transition.  \\

	The combination of X-ray diffraction, M\"ossbauer and NMR data unambiguously identify the low temperature state of the as grown (quenched from $960^{\circ}$ C) sample as being non-magnetic and also having a collapsed tetragonal unit cell that is remarkably similar to what has been found for Sn-grown CaFe$_2$As$_2$ under hydrostatic pressure.  For that matter, the evolution of the temperature dependent resistivity as well as magnetic susceptibility are both qualitatively similar to the evolutions found when pressure is applied as hydrostatically as possible, i.e. with He as a pressure medium. \cite{yuw09a}  At this point, having not only created a $T^* - T_a$ phase diagram that looks a lot like the $T - P$ phase diagram (for $T_a > 400$$^{\circ}$ C), Fig. \ref{F5},  but also having clearly identified the phases associated with this phase diagram, it is appropriate to investigate the possible physical origin, or mechanism, for this apparent similarity between pressure applied to a Sn-grown crystal of CaFe$_2$As$_2$ and annealing of FeAs grown crystals.\\

	A starting point for this search for a mechanism can be found in a subset of the observations made above.  The as grown crystals from FeAs solution are far more brittle than either the Sn-grown crystals or the FeAs grown crystals after a 400$^{\circ}$ C anneal.  This qualitative observation hints at some higher concentration of defects in the as grown crystals that lead to embrittlement.  In addition, both the M\"ossbauer and NMR measurements find broader line shapes associated with the spectra from the as grown samples, indicating that there may be a greater degree of disorder in them than in the crystal annealed at 400$^{\circ}$ C.  \\

In order to examine the distribution of defects at a nano-scale level, TEM measurements were carried out on both as grown, quenched from 960$^{\circ}$ C, samples as well as samples that had been annealed at 400$^{\circ}$ C for a week.  The as-grown sample (Fig. \ref{F12}(a)) shows a pervasive tweed-like pattern with $\sim 40$ nm separation of features.  The selected area diffraction pattern, (inset, Fig. \ref{F12}(a)) shows only the $[0,0,1]$ zone axis pattern consistent with the CaFe$_2$As$_2$ compound and no additional reflections or streaking indicative of a superlattice or a highly defective (intercalated) lattice.  These very long but thin features are orthogonal and are approximately parallel to the $\{h,0,0\}$ planes as best can be determined in this orientation of the sample.  The thinnest regions of the sample did not exhibit these features, consistent with the ease at which these samples were damaged by ion milling.  (Milling above 3 keV and not cooling with liquid nitrogen resulted in significant milling artifacts.)  The thickness of the foils where these features are present and the lattice strain they cause, prevent atomic resolution imaging at this point.  However, tilting experiments and imaging with the principle reflections did reveal the two-dimensional nature of these thin lamellae.  These features were consistent throughout all the thin area of the sample although in some regions one variant may dominate over the other and in some regions interpenetrating lamellae were observed as shown in Fig. \ref{F12}(a).  Occasional dislocations were observed, but did not dominate the microstructure. \\

The sample annealed at 400$^{\circ}$ C for one week appears completely different.  Here we observed a very smooth contrast across the thin region when tilting and uniformly distributed small lenticular precipitates about 25 to 100 nm in width and with a length to width aspect about 5:1 (Fig.  \ref{F12}(b)). These precipitates are also fairly uniformly separated, $\sim 500$ to 1,000 nm, and have their long axis parallel to the $\{h,0,0\}$, as was observed in the tweed pattern of the 960$^{\circ}$ C quenched sample.  Dislocations in the matrix are commonly observed to emanate from the interface between the precipitates and the matrix typically near the ends of the precipitate where stresses would be higher if there are differences in coefficients of thermal expansion.  The SADP is nearly identical to the as grown sample but here the precipitates are large enough for diffraction analysis.  The convergent beam electron diffraction (CBED) pattern, RHS inset Fig. \ref{F12}(b), produces disks rather than spots due to how the pattern is formed, but it is clear that the pattern is identical to that of the matrix.  The fact that the SADP did not show any splitting of spots when including the precipitates and the matrix in the same sample area would suggest that in this orientation the two lattices are nearly coherent.  EDS also indicated that the precipitate phase does not contain Ca.  Efforts to identify the precise chemistry and structure of this second phase are on-going but it should be noted that the basal plane dimensions of the CaFe$_2$As$_2$ and the tetragonal AsFe$_2$ (space group $P4/nmm$, \# 129) are within 5 \% and have very similar $[0,0,1]$ diffraction patterns.  A gross estimate of the impurity phase area in less than 5 \% of the total sample, giving a gross idea of how much extra Fe and As is trapped in the sample when it is initially quenched from 960$^{\circ}$ C. \\

The similarity in the orientation and lack of distinguishing features in the diffraction between the as grown samples and the samples that were annealed suggests that there is a similarity in chemistry/structure between the tweed strain fields and coarser precipitates in these two samples and the difference is simply one of length scale.  An epitaxial relationship would lower the energy barrier for nucleation and allow a second phase to form more readily if thermodynamically stable.  Annealing at moderate temperatures but within a two-phase field, would promote growth of the second phase to reduce the excess energy due to interfaces (i.e., Oswald ripening).  The observations here are consistent with an increase in the width of formation of CaFe$_2$As$_2$ with respect to excess As and Fe at elevated temperatures which decreases monotonically with temperatures below 960$^{\circ}$ C.\\

Such a temperature-dependent solid solubility of excess Fe and As leads to the following, plausible, scenario. When the FeAs flux grown crystal is first quenched, there is little time for the excess As and Fe to come out of solution.  In these single crystals, the grain dimensions (often mm to cm) are simply too large for diffusion to allow for the expulsion of these species to the grain boundaries. Energetically, it appears easiest to exsolve excess As and Fe epitaxially along the $\{h,0,0\}$ planes.  Differences in their unit cell size as well as their coefficients of thermal expansion (CTE) can lead to significant stresses at the interfaces between the CaFe$_2$As$_2$ majority phase and the finely dispersed Fe/As based second phase. If, as Fig. \ref{F12}a would suggest, domains of CaFe$_2$As$_2$, about 40 nm on a side, are surrounded by nearly coherent second phase resulting in a significant volume fraction of interfaces or regions strained by interfaces, then the magnitude of the stress would be dependent on the volume fraction of the CaFe$_2$As$_2$ in these strained regions.\\

If there is a temperature dependence of the solubility of the excess As and Fe, then quenching from lower temperatures would result in a smaller fraction of finely dispersed second phase, the remaining excess As and Fe being sequestered in larger precipitates whose insignificant surface to volume ratio would have little impact on the matrix (as is the case in Fig. \ref{F12}b).  This smaller amount of finely dispersed precipitate would lead to a smaller average strain (or pressure) on the sample, leading to an effective correlation between $P$ and $T_a$.  Since the initial quench of the large crystals from the flux essentially locks in the excess As and Fe, the subsequent processing history determines the size and distribution of the second phase and thereby determines the amount of strain in the sample.\\

Further exploration of this hypothesis requires confirmation of the second phase crystal structure and its chemistry and determination of the CTE and bulk moduli of these two phases.  In addition, returning to the initial motivations for this study: given that  annealing of as grown samples seems to lead to small changes in the transition temperatures of BaFe$_2$As$_2$ based compounds \cite{rot10a,gof10a} and given that CaFe$_2$As$_2$ is much more pressure / strain sensitive than BaFe$_2$As$_2$, it is worth exploring the implications of our current findings.  If we speculate that a similar width of formation exists in BaFe$_2$As$_2$ (or for that matter SrFe$_2$As$_2$) and that low temperature annealing can lead to similar effects as those we present here, then, based on the existing $P - T$ phase diagrams, \cite{col09a} an effective pressure of $\sim 0.4$ GPa on BaFe$_2$As$_2$ or SrFe$_2$As$_2$ would only lead to shifts in $T^*$ of a few K, consistent with what has been observed. \cite{rot10a,gof10a}  Based on this analysis, TEM measurements on as grown and annealed crystals of BaFe$_2$As$_2$, as well as SrFe$_2$As$_2$, to check for similar, annealing temperature dependent microstructure are in order. 

\section{Conclusions}

\begin{figure}
\begin{center}
\includegraphics[angle=0,width=100mm]{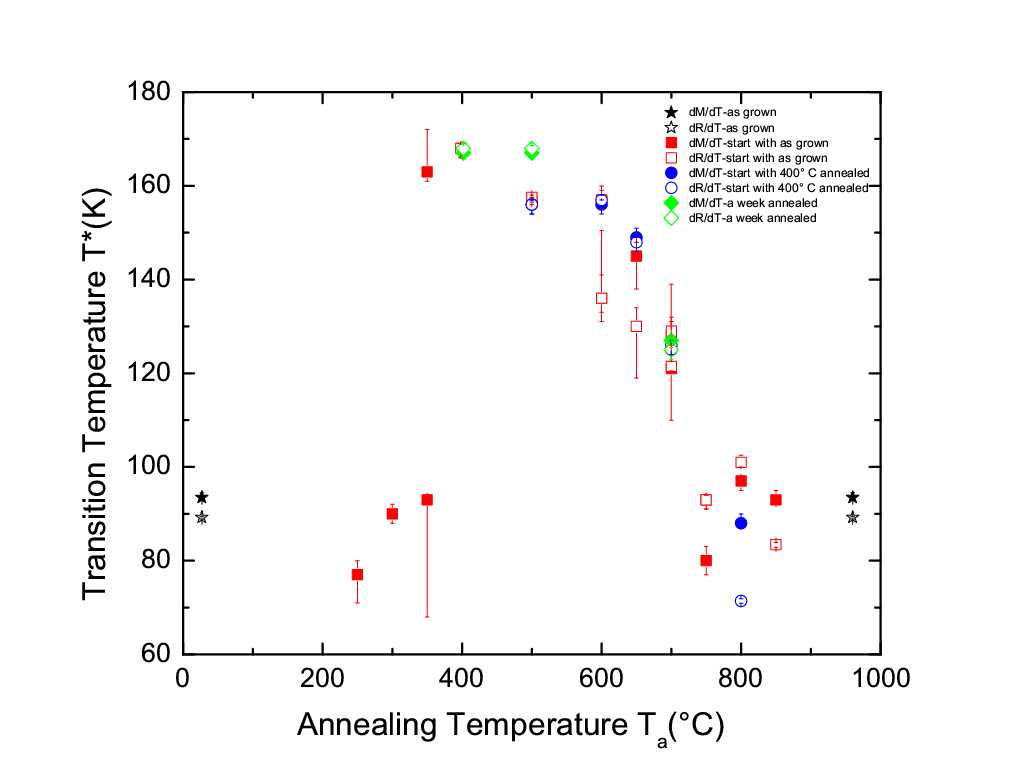}
\includegraphics[angle=0,width=100mm]{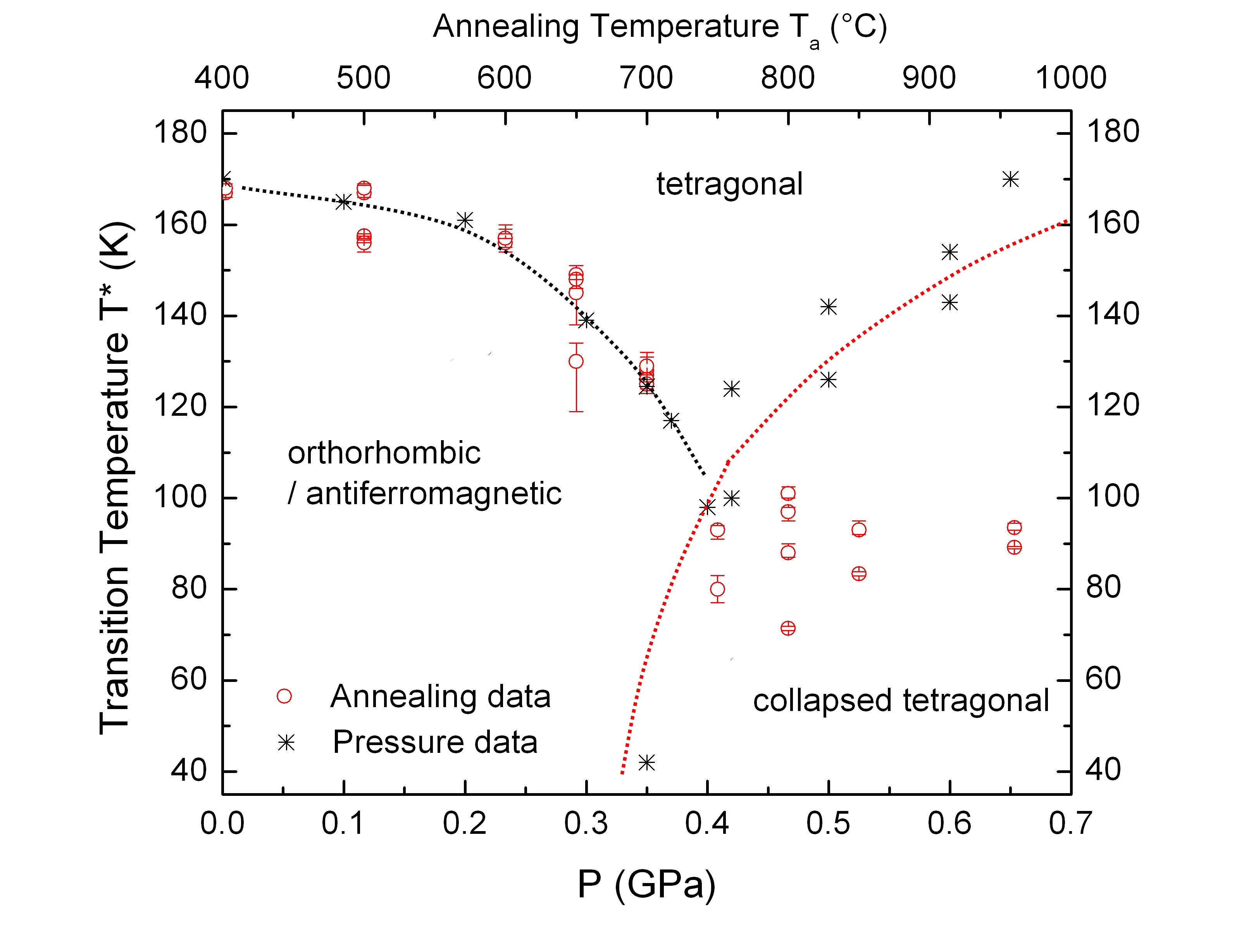}
\end{center}
\caption{\label{F5} (Color online)  (a) Transition temperature ($T^*$) – annealing temperature ($T_a$) phase diagram.  Open symbols are inferred from resistivity data and filled symbols are inferred from susceptibility data.  Stars are the as grown samples  (quenched from 960$^{\circ}$ C data and are also shown as 20$^{\circ}$ C anneals); squares are as grown samples that have been annealed for 24 hours at $T_a$ and quenched to room temperature; circles are as grown samples that were first annealed for a week at 400$^{\circ}$ C and then annealed for 24 hours at $T_a$ and quenched to room temperature; diamonds are as grown samples that have been annealed for a week as whole, unopened batches at $T_a$.  (b) $T^*$ as a function of pressure from Ref. \onlinecite{yuw09a} and $T^*$ as a function of $T_a$ for $400$$^{\circ}$  C$ \leq T_a \leq 960$$^{\circ}$ C.  
 }
\end{figure}

\begin{figure}
\begin{center}
\includegraphics[angle=0,width=100mm]{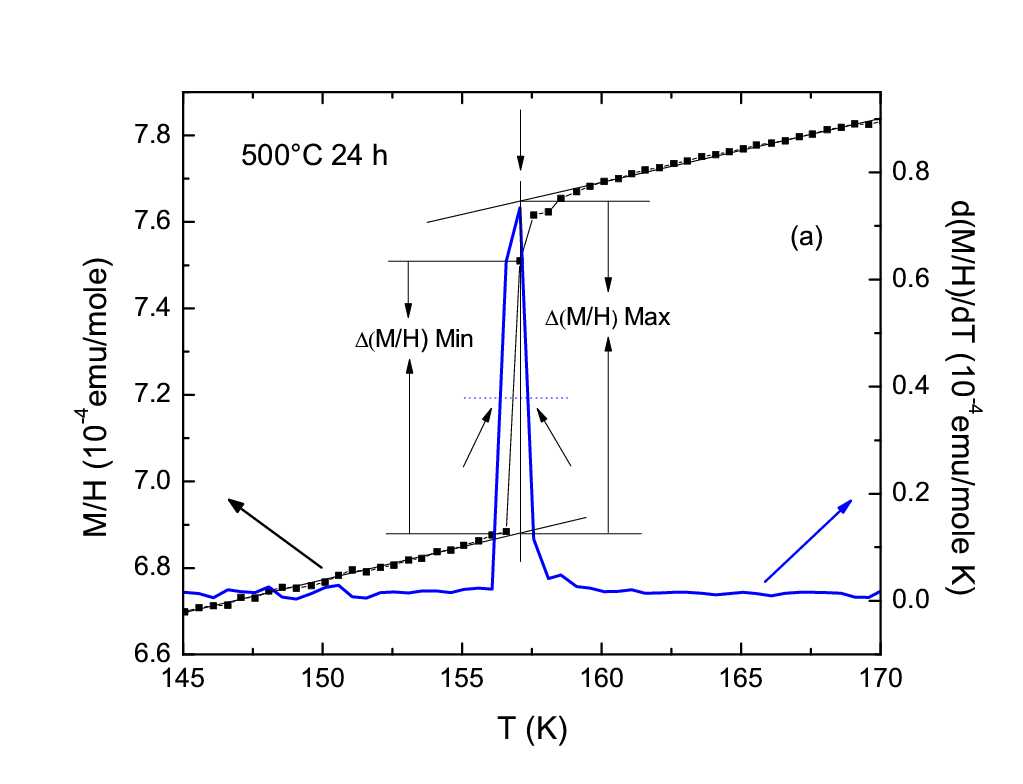}
\includegraphics[angle=0,width=100mm]{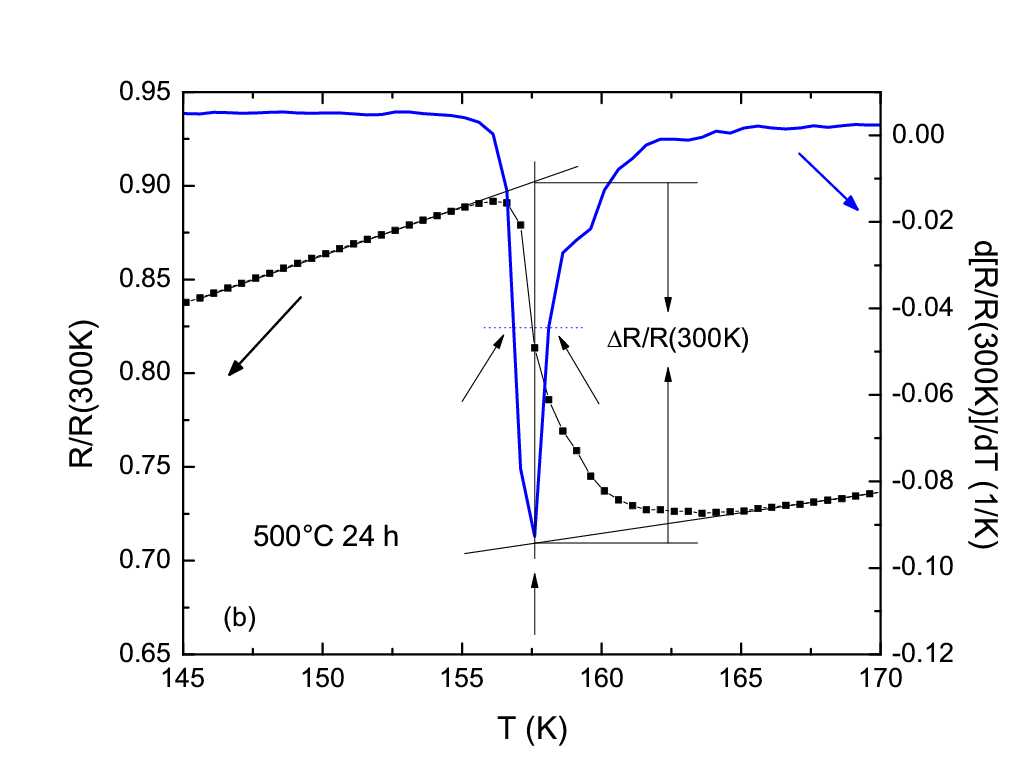}
\end{center}
\caption{\label{F6} (Color online) Temperature dependent (a) magnetic susceptibility and (b) normalized electrical resistivity and their respective temperature derivatives for an as grown crystal annealed at 500$^{\circ}$ C for 24 hours. $T^*$ is inferred from the temperature of the extremum in the derivative.  Full width at half maximum of derivative is used to define error bars shown in Fig. \ref{F5}.  Values of the jump in magnetic susceptibility (maximum and minimum values) and resistivity are inferred as shown.  }
\end{figure}

	We have found a remarkably large response of the transition temperature of  CaFe$_2$As$_2$ single crystals grown out of excess FeAs to annealing / quenching temperature.  Whereas crystals that are annealed at 400$^{\circ}$ C exhibit a first order phase transition from a high temperature tetragonal to a low temperature orthorhombic and antiferromagnetic state near 170 K (similar to what has been found in the original Sn grown single crystals \cite{nin08a,can09a}), crystals that have been quenched from 960$^{\circ}$ C exhibit a transition from a high temperature tetragonal phase to a low temperature, non-magnetic, collapsed tetragonal phase below 100 K.  We have been able to demonstrate that the transition temperature can be reduced in a monotonic fashion by varying the annealing / quenching temperature from 400 to 850$^{\circ}$ C with the low temperature state remaining antiferromagnetic for transition temperatures larger than 100 K and becoming collapsed tetragonal / non-magnetic for transition temperatures below 90 K.  This suppression of the orthorhombic / antiferromagnetic phase transition and its ultimate replacement with the collapsed tetragonal / non-magnetic phase is similar to what has been observed for Sn-grown single crystals of  CaFe$_2$As$_2$ under hydrostatic pressure. \cite{yuw09a}  This similarity is summarized in Fig. \ref{F5}b.  

	TEM studies of the as grown (quenched from 960$^{\circ}$ C) and annealed crystals indicate that there is a temperature dependent, width of formation of  CaFe$_2$As$_2$ with a decreasing amount of excess Fe and As being soluble in the single crystal at lower annealing temperatures.  On one extreme, samples quenched from 960$^{\circ}$ C have, finely divided, strain structure with characteristic length scales and spacings of less than 50 nm.  On the other extreme, samples annealed at 400$^{\circ}$ C have clearly identifiable, Ca free, crystalline inclusions with dimensions of $\sim 70 \times 500$ nm$^2$ that are separated by 500 to 1,000 nm.  These images make it clear that when the sample is quenched from 960$^{\circ}$ C it is possible to think of some average, near uniform strain throughout the sample associated with the overlapping strain fields of this fine precipitate.  It is this strain that appears to be giving rise to the dramatic suppression of $T^*$, in, apparently, a similar manner that hydrostatic pressures of $\sim 0.4$ GPa can.

	Finally, it is worth noting that, as was the case with hydrostatic pressure applied by helium pressure medium, \cite{yuw09a} there is no indication of superconductivity, even in trace amounts.  This would be consistent with the idea that superconductivity comes from the poorly defined strains associated with the onset of the collapsed phase in a confined (solidified pressure medium) volume.  Given that these measurements are at ambient pressure, this is not the case. By the same token, though, given that we can essentially mimic the key effects of hydrostatic pressure via variation of $T_a$, the collapsed tetragonal phase, as well as the orthorhombic / antiferromagnetic phase with intermediate $T^*$ values, is open to measurements at ambient pressure.  This should make detailed thermodynamic, microscopic and spectroscopic measurements across the phase space outlined in Fig. \ref{F5}b much more accessible.

\begin{acknowledgments}
We thank D. Robinson for his excellent technical support of the X-ray diffraction study. This work was supported by the U.S. Department of Energy, Office of Basic Energy Science, Division of Materials Sciences and Engineering.  Ames Laboratory is operated for the U.S. Department of Energy by Iowa State University under Contract No. DE-AC02-07CH11358. Work at  McGill University was supported by
grants from the Natural Sciences and Engineering Research Council of Canada and Fonds Qu\'eb\'ecois de la Recherche sur la Nature et les Technologies.
\end{acknowledgments}

\clearpage

\clearpage

\clearpage

\clearpage

\clearpage

\clearpage

\clearpage

\begin{figure}
\begin{center}
\includegraphics[angle=0,width=120mm]{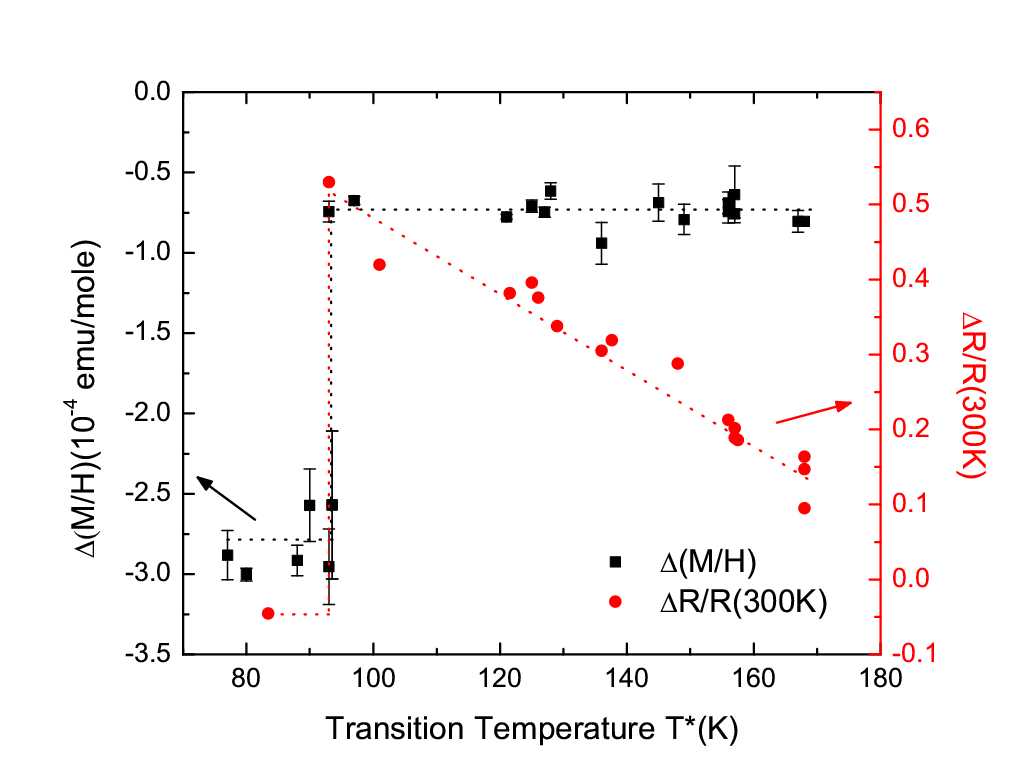}
\end{center}
\caption{\label{F7} (Color online)  Size of jump in susceptibility and normalized resistivity as a function of transition temperature for FeAs grown CaFe$_2$As$_2$ crystals annealed at temperatures shown in Figures \ref{F1} -  \ref{F4}.  Room temperature resistivity of samples with $T^*$ values of $\sim 90$, 130 and 170 K all fall within the $3.75 \pm 0.75$ m$\Omega$ cm range.
 }
\end{figure}

\clearpage

\begin{figure}
\begin{center}
\includegraphics[angle=0,width=70mm]{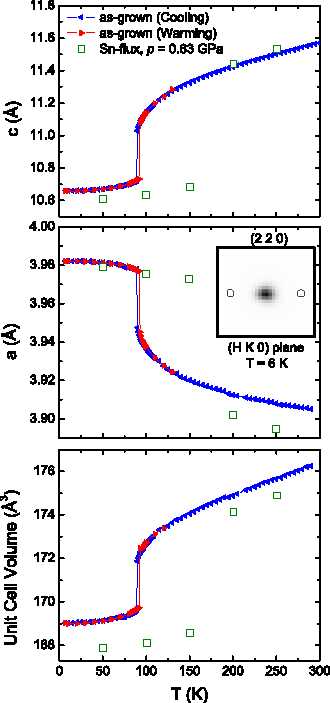}
\end{center}
\caption{\label{F8} (Color online)  Values for (a) the $c$-lattice paramter, (b) $a$-lattice parameter and (c) unit cell volume as a function of temperature for an unannealed FeAs flux-grown CaFe$_2$As$_2$ sample determined from high-energy X-ray diffraction measurements.  The open squares denote the results of measurements performed on a polycrystalline sample under applied hydrostatic pressure of 0.63 GPa from reference \onlinecite{kre08a}.  The inset to the middle panel is the image of the $(2~2~0)$ diffraction peak taken from the two-dimensional X-ray detector as described in the text.  Note the absence of any splitting that would signal a transition to an orthorhombic phase [the two open circles illustrate the expected distance between split Bragg peaks due to the "usual" orthorhombic distortion ]. }
\end{figure}

\clearpage

\begin{figure}
\begin{center}
\includegraphics[angle=0,width=120mm]{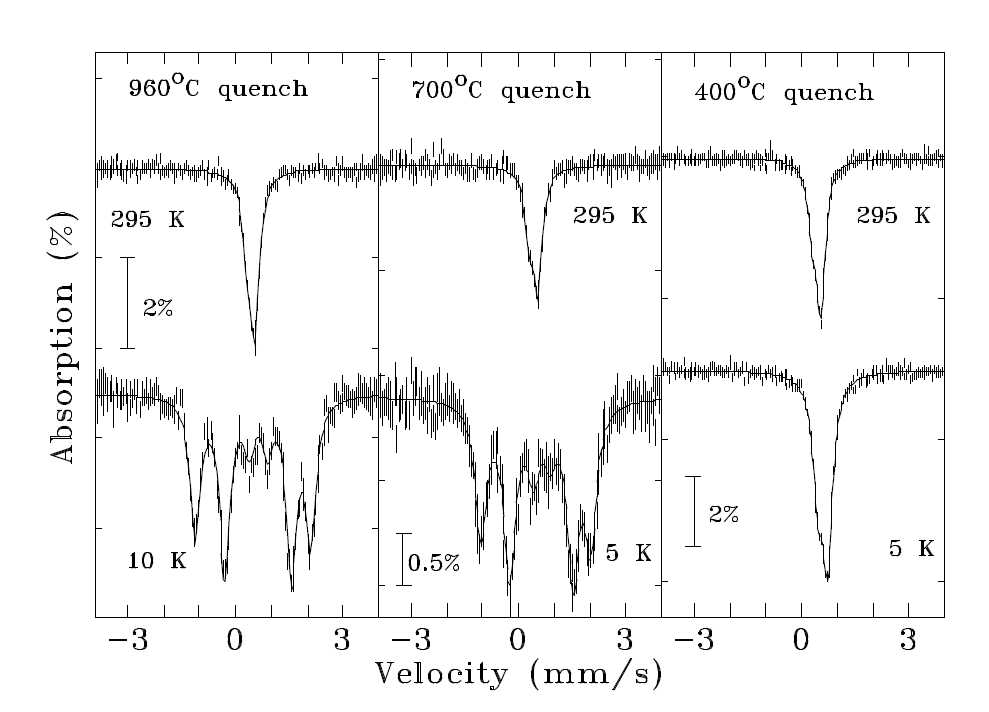}
\end{center}
\caption{\label{F9}  $^{57}$Fe M\"ossbauer spectra of ab-plane single crystal mosaic of samples (from left) annealed at 400$^{\circ}$ C, annealed at 700$^{\circ}$ C and as grown (quenched from 960$^{\circ}$ C). In each case the upper spectrum was taken at 295 K whereas the lower spectrum was taken at 10 K or 5 K. Only the as grown sample shows no magnetic ordering at base temperature. The solid lines are fits as described in the text. }
\end{figure}

\clearpage

\begin{figure}
\begin{center}
\includegraphics[angle=0,width=140mm]{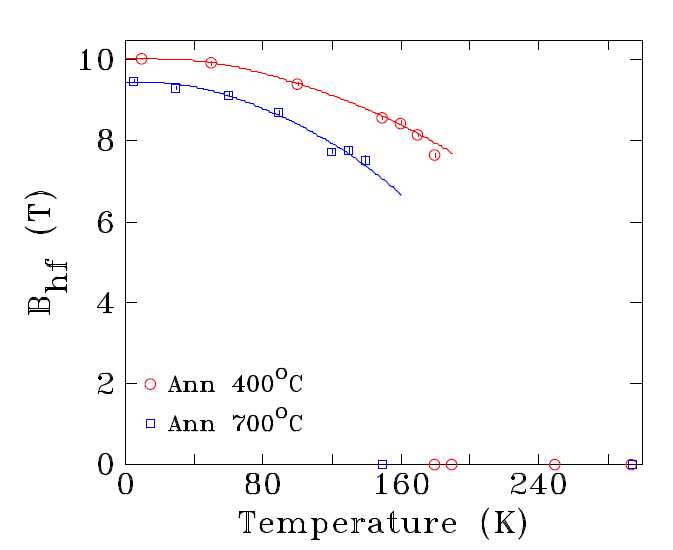}
\end{center}
\caption{\label{F10} (Color online)   Temperature dependence of the magnetic hyperfine fields ($B_{hf}$) for the two annealed samples shown in Figure \ref{F9}. The solid lines are fits to Brillouin functions (as described in the text) used to estimate the N\'eel temperatures of the low temperature antiferromagnetic orthorhombic form of each sample. In both cases the samples undergo a first order structural transition on warming effectively truncating the $B_{hf}(T)$ data.
}
\end{figure}

\clearpage

\begin{figure}
\begin{center}
\includegraphics[angle=0,width=120mm]{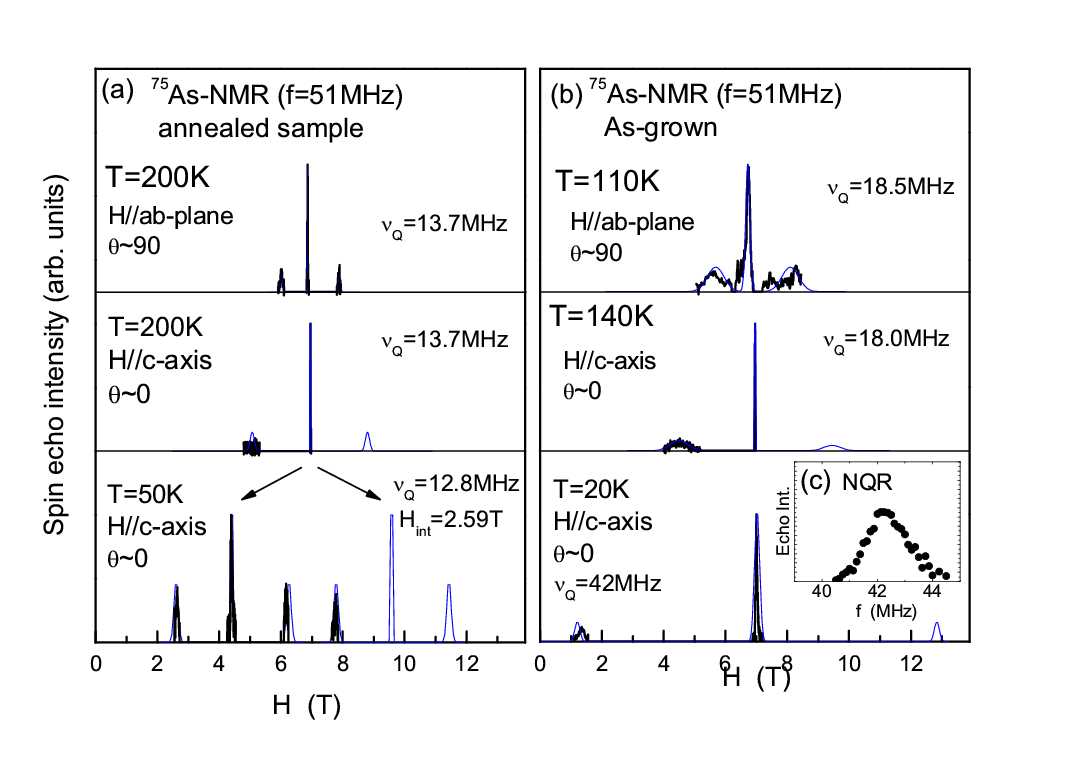}
\end{center}
\caption{\label{F11} (Color online)  $^{75}$As-NMR spectra measured at $f = 51$ MHz for  (a) $400 ^{\circ}$ C annealed CaFe$_2$As$_2$ crystal and (b) for the as-grown CaFe$_2$As$_2$ crystal. Black and blue lines are observed and simulated spectra, respectively. Expected lines above 9 T are not measured due to the limited maximum magnetic field for our SC magnet. (c)  $^{75}$As NQR spectrum at $T = 4.2$ K and $H = 0$ T. }
\end{figure}

\clearpage

\begin{figure}
\begin{center}
\includegraphics[angle=0,width=90mm]{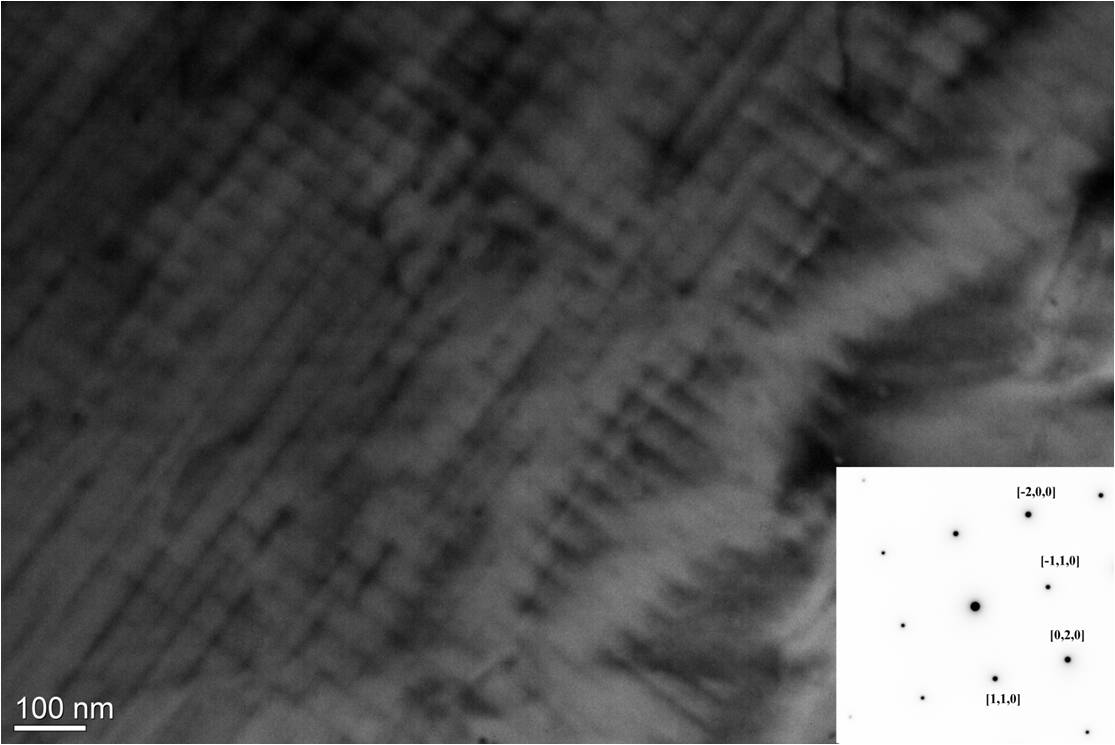}
\includegraphics[angle=0,width=90mm]{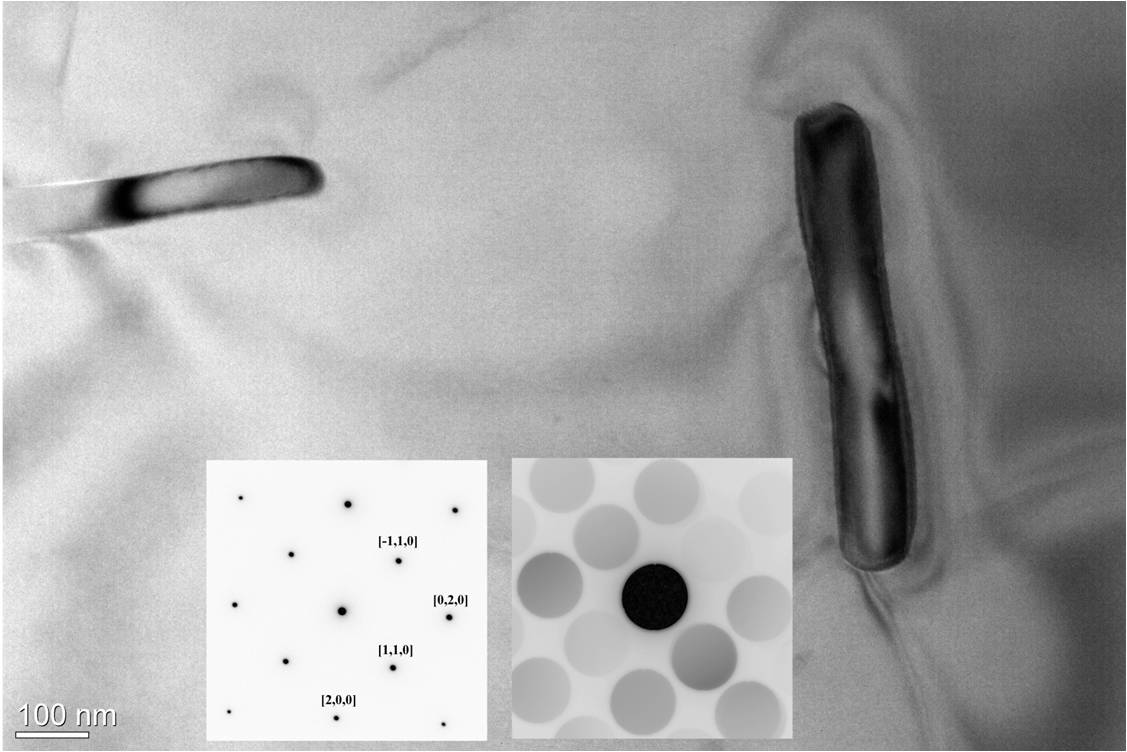}
\end{center}
\caption{\label{F12} (a) TEM micrograph of the as-grown and quenched sample.  The fine tweed-like pattern is due to thin platelets of a second phase which are coherent with the matrix and parallel the $\{h,0,0\}$ planes.  The lower right-hand side inset is the SADP from this image area showing expected lattice reflections for a $[0,0,1]$  zone axis for the 122 compound. Note the absence of streaking or extra reflections as discussed in the text. (b) TEM micrograph of the sample annealed at 400$^{\circ}$ C for 1 week at the same magnification and zone axis orientation as the as-grown sample above.  Note that the rectangular precipitates are oriented along the same $\{h,0,0\}$ planes as the finer features above.  The left-hand inset shows the SADP of the matrix and precipitates while the right-hand inset is a CBED of only the precipitate phase showing nearly the same orientation and d-spacing as the matrix 122 phase.   }
\end{figure}

\end{document}